\documentstyle{ichep}
\input psfig
\newcommand{\ttbar}{$t \bar t$ }
\newcommand{\ppbar}{$p \bar p$}
\newcommand{\Wjet}{$W+{\rm jets}$}
\newcommand{\ejet}{$e+{\rm jets}$}
\newcommand{\mujet}{$\mu+{\rm jets}$}
\def\Etmisscal{\hbox{$\rlap{\kern0.25em/}E_T^{cal}$}}
\def\Etmiss{\hbox{$\rlap{\kern0.25em/}E_T$}}
\begin{document}
\title{Search for top in lepton + jets in D\O\ using a topological tag}
\author{Serban Protopopescu}

\affil{Brookhaven National Laboratory,\\
Upton, New York 11973, USA}

\collab{On behalf of the D\O\ collaboration}

\abstract{We have searched for production of \ttbar\ pairs in
\ppbar\ interactions at 1.8 TeV center-of-mass energy at the
FNAL Tevatron collider. The search assumes
standard model decay for top quark into W + b quark. We observe
in \ejet\ and \mujet\ final states
a small, not statistically significant, excess above the background
estimated by two different methods. The results presented are preliminary.}

\twocolumn[\maketitle]

\section{Introduction}

    Assuming the Standard Model decay for top quarks into a W and a b
quark, events with \ttbar\ pairs will have  one isolated
high $p_T$ e or $\mu$ accompanied by multiple jets
30\% of the time. These channels
yield considerably more events than those with 2 isolated
leptons in the final state, presented in the preceding contribution
to these proceedings by S. Wimpenny \cite{Wimpenny}
, but suffer from larger backgrounds
from the channel W+ multi-jets. In this paper we discuss two different
methods used to estimate whether a significant \ttbar\  production
can be observed in the D\O\ data over the \Wjet\ background.
A previous analysis of these data, searching for top in the mass
region 90 to 140 GeV in single and di-lepton final states,
 has been published \cite{PRL}
and gave a top mass limit of 131 GeV (95\% CL). The present analysis
is optimized for masses above 130 GeV and should be considered preliminary.
The two methods used yield a small, but not statistically significant, excess.
\section{Event selection}
    The D\O\ detector is described in detail in \cite{D0} and in these
proceedings by \cite{Wimpenny}, so it will not be discussed here.
    The events for lepton + jets analysis are obtained with 3 triggers:
\begin{itemize}
\item One electromagnetic shower with
transverse energy ($E_T$) $>15$ GeV, one hadronic
jet with $E_T>10$ GeV and calorimeter missing $E_T~ (\Etmisscal)~ >10$ GeV
\item One electromagnetic shower with $E_T>20$ GeV and $\Etmisscal>20$ GeV
\item One $\mu $ with $p_T>8$ GeV and one hadronic jet with $E_T>15$
GeV
\end{itemize}
Hadronic jets are defined as energy in
a cone of $R=\sqrt{\Delta \phi ^2 + \Delta \eta ^2}=0.5$ where $\phi$
is the azimuthal angle in radians and $\eta$ is the pseudo-rapidity.

After reconstruction the
events are selected further by making stricter requirements. For
the \ejet\ sample selection requirements on the
electron are based on
shower shape, cluster isolation, shower match with a reconstructed
track and track ionization cuts to remove conversions.
 Electrons are required to have
$E_T^e>20$ GeV and
$|\eta ^e|<2.0$. Additional requirements on the \ejet\ sample
are $\Etmisscal >25$ GeV, $E_T^{jet}>15$ GeV and $|\eta_{jet}|<2.0$.
\begin{figure}
\vspace*{3pc}
\centerline{\psfig{figure=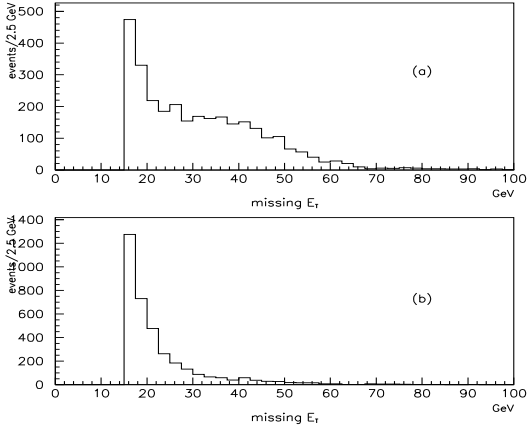,width=3.0in,height=2.5in}}
\caption{(a) $\Etmisscal$ distribution for events satisfying
all electron selection criteria, (b) $\Etmisscal$ distribution for
events satisfying the electron triggers but failing offline selection
criteria.}
\label{fig-met}
\end{figure}
Figure  \ref{fig-met}a (\ref{fig-met}b) shows the $\Etmiss$ distribution
for the selected events
which pass (fail) the electron cuts. We will refer to the events that
satisfied the triggers
but failed the offline electron cuts and have
$\Etmisscal<25$ GeV as the QCD sample.
This sample is used to estimate the non-W background in the
\ejet\ sample and to test the validity of jet multiplicity scaling
in section \ref{jetmult}.

Selection requirements on the $\mu$ are made to remove cosmic rays
(by drift chamber timing and no other collinear $\mu$-track),
isolation cuts to remove $\mu$'s from decays other than W's
( minimum separation between $\mu$ and jets with
$E_T>8$ GeV of $\Delta R>0.5$ and core cone isolation
energy $<5$ GeV in annular
cone $0.2<\Delta R<0.4$ around $\mu$ track) plus calorimeter dE/dx confirmation
to reduce accidentals. The $\mu$ is required to have
$p_T^{\mu}>15$ GeV and $|\eta_{\mu}|<1.7$. In addition to the $\mu$ the
\mujet\ sample is required to have at least 1 jet with $E_T>15$ GeV
and $|\eta_{jet}|<15$ GeV,  $\Etmiss >20$ GeV, and $\Etmisscal>20$ GeV.
The trigger and selection requirements on the \mujet\ sample produces
a bias for $\leq 2$ jets which needs to be corrected when added to
\ejet\ for the jet multiplicity analysis in section \ref{jetmult}.

To avoid overlap with the analysis of events in which jets are tagged
as containing heavy quarks by the observation of a non-isolated $\mu$,
discussed in the contribution to these proceedings by R. Raja \cite{Raja},
$\mu$-tagged events have been excluded from the sample and their loss accounted
for in the acceptance calculation.

\section{Jet Multiplicity Analysis}
\label{jetmult}
    The number of events as function of jet multiplicity and the
estimated non-W background to \ejet\ and \mujet\ channels are
given in table \ref{tab-jetmult}. The non-W background for \ejet\
is mostly
from misidentified electrons. In \mujet\ events the background comes from
two sources: Z + jets where one of the $\mu$ is not detected and QCD
processes
where the $\mu$ from a decay other than $W$ satisfies the isolation criteria,
the two backgrounds are roughly comparable.
\begin{table}
\Table{|l||c|c||c|c|}{
$jets$ & \ejet\  & non-$W$ bckg. &
\mujet\  &  non-$W$ bckg.    \\
\hline
$\geq 1$ & 1383  & $94. \pm 6.$  & 303  & $48. \pm 8.$  \\
$\geq 2$ & 243 &  $26 \pm 3.$ & 92 &  $16. \pm 3.$  \\
$\geq 3$ & 35 & $5.7 \pm 0.9$ & 22 & $4.8 \pm 0.8$  \\
$\geq 4$ & 6 &  $1.1 \pm 0.3$ & 6 &  $1.3 \pm 0.3$  \\
$\geq 5$ & 2 &  $0.3 \pm 0.1$ & 1 &  $0.3 \pm 0.1$  \\ \hline
}
\caption{Selected e and $\mu$ + jets events}
\label{tab-jetmult}
\end{table}

\begin{figure}
\vspace*{4pc}
\centerline{\psfig{figure=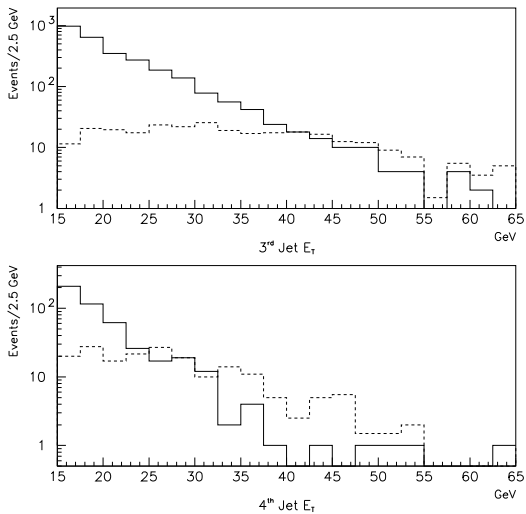,width=3.0in,height=3.0in}}
\caption{$E_T$ distributions of $3^{rd}$ and $4^{th}$ jet
from $W+ \ge 3$ jets Montecarlo
(solid) and \ttbar\ (dotted) with top mass 160 GeV, normalized
to 500 $pb^{-1}$.}
\label{fig-jetet}
\end{figure}

\begin{figure}
\vspace*{5pc}
\centerline{\psfig{figure=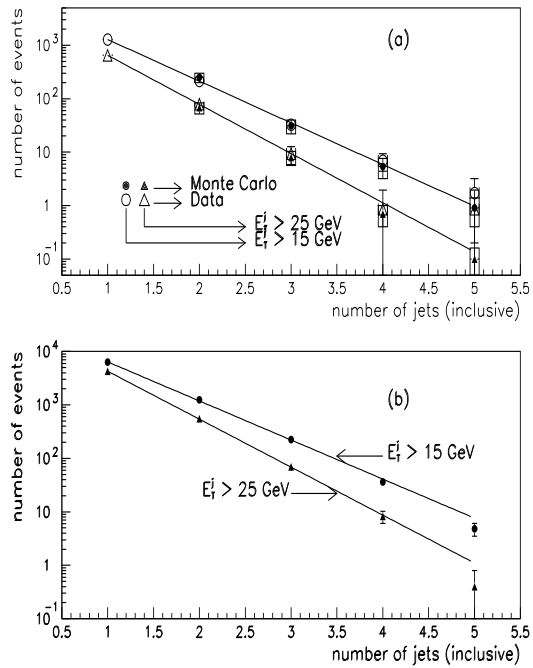,width=3.0in,height=4.0in}}
\caption{Number of events as function of jet multiplicity
(a) \ejet\ data after non-W background subtraction and
VECBOS predictions, boxes
indicate uncertainty in prediction, (b) QCD sample after
correcting for increased probability for triggering with
increasing multiplicity.}
\label{fig-ejet}
\end{figure}

As shown on Fig. \ref{fig-ejet}(a)
the \ejet\ data after background subtraction are fit well by a scaling
law (as expected by \cite{Berends}):
$$W +  n~ jets=\frac{[W+ (n-1)~jets] \cdot [W+ (n-1)~jets)]}
{[W+ (n-2)~jets)]}$$
This law seems to be satisfied for any $E_T^{jet}$ threshold used for
counting jets. The predictions from the VECBOS \cite{VECBOS} Montecarlo
agree well with the data after ISAJET \cite{ISAJET} is used for hadronization
and GEANT \cite{GEANT} for simulating the D\O\ detector.
As \ttbar\ events in channels with one isolated lepton
are expected to have high multiplicities (2 jets from b's, 2 jets from
hadronic W decay plus additional jets from initial and final state
radiation) any noticeable \ttbar\ production should appear as an excess
of events at high multiplicity, in particular for 4 or more jets
and for high $E_T$ threshold. As shown in Fig. \ref{fig-jetet} the 4th
jet (ordered in $E_T$) from high mass top is expected to be considerably
more energetic than for \Wjet\ but no significant violation of the
scaling law is observed in Fig. \ref{fig-ejet}(a).
To find how much \ttbar\ production can be accomodated
and still satisfy the scaling law we can solve using the number of events
with 2,3 and 4 jets for the number of W and top events:
$$N_4-a_4 \cdot N_t=(N_3-a_3 \cdot N_t)^2/(N_2-a_2 \cdot N_t)$$
where $N_i$ is observed number of lepton + jets events with $\geq$i jets,
$N_t$ the total number of top events and $a_i$
the corresponding fraction at each multiplicity. The results are given
in table \ref{tab-jetfit}, of particular interest is the total number of top
events $N_t=5.7 \pm 10.0$. The numbers for $\geq 1$ jets are predicted
from the numbers obtained using the events with 2,3 and 4 jets.

\begin{table}
\Table{|l||c|c||c|c|}{
 jets& Data & non $W$ & \multicolumn{2}{|c|}{calculated}\\
  ~  &  ~   & backg.  & \Wjet\  & \ttbar\ \\
\hline
$\geq 1$ & 1686  & $142. \pm 20.0$ & $1656 \pm 96$ & $5.7 \pm 10.0$ \\
$\geq 2$ & 335 &  $50.6 \pm 7.0$ & $278.8 \pm 17 $ & $5.6 \pm 9.9$ \\
$\geq 3$ & 57 & $10.3 \pm 1.5$ & $42.2 \pm 7.1$ & $4.5 \pm 8.2$ \\
$\geq 4$ & 12 &  $2.4 \pm 0.4$ & $6.6 \pm 3.8$ & $3.0 \pm 4.6$ \\ \hline
}
\caption{Estimated numbers of \Wjet\ and \ttbar\ events}
\label{tab-jetfit}
\end{table}

To check the validity of the scaling
law in a case with no \ttbar\ events, we fitted
the QCD sample to a simple exponential. In Fig. \ref{fig-ejet}(b) we show
the jet multiplicities for that sample (after correcting for the higher
probability as a function of
increasing multiplicity for having a jet fake an electron).
The QCD data is clearly well fitted. We also fitted the predictions of the
VECBOS Montecarlo. All the fits give similar values for the ratio of
$[W+3~{\rm jets}]/[W+2~{\rm jets}]$:
$.151 \pm .020$ for the data, $.160 \pm .013$ for VECBOS Montecarlo,
$.187 \pm .004$ for the QCD sample. The fit to the VECBOS Montecarlo
tends to underestimate the predicted number of 4 or more jet events
by 15\%
while the QCD fit overestimates by the number of 4 jet events by 12\%.
To account for these
deviations we estimate a 20\% systematic error in the validity
of the scaling law. The fraction ($a_i$) of top events
 expected at each multiplicity
is obtained from Montecarlo; to estimate the uncertainty in this procedure
we compared 2 different Montecarlos (ISAJET and HERWIG \cite{HERWIG}).
The differences between them were at the level of 10\%. There is an
additional 15\% uncertainty in the top acceptance from the uncertainty
in the jet energy scale. From the number of top events and the
calculated acceptance times branching ratio
we can estimate a cross section for \ttbar\ production:
$$\sigma_{t \bar t}=6.4 \pm 9.8~ {\rm (stat.)}\pm 4.0~{\rm (sys.)}~pb $$
the sytematic error includes a 12\% uncertainty in the luminosity.

This cross section is for masses between 160 and 180 GeV. As the acceptance
falls moderately below 160 GeV it should be increased by 25\% around 140 GeV.

\section{${\cal A},~H_T$ Analysis}

\begin{figure}
\vspace*{4pc}
\centerline{\psfig{figure=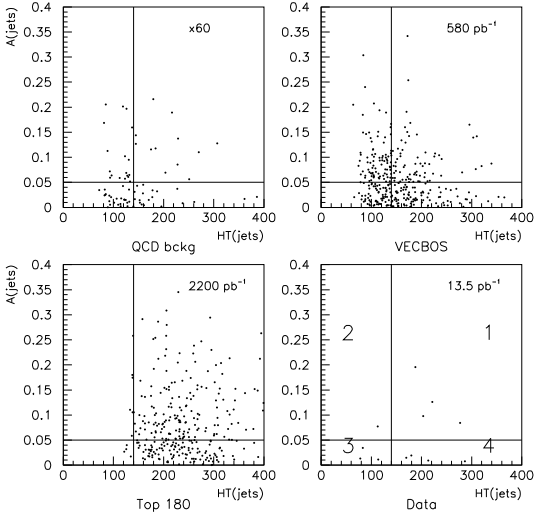,width=3.0in,height=3.0in}}
\caption{${\cal A}$ vs $H_T$ distributions for events with 4
or more jets: (a) QCD sample, (b) VECBOS Montecarlo,
(c) \ttbar\ events, top mass 180 GeV, (d) data, the
quadrants used in the fit are numbered, see text.}
\label{fig-aht}
\end{figure}

\begin{figure}
\vspace*{4pc}
\centerline{\psfig{figure=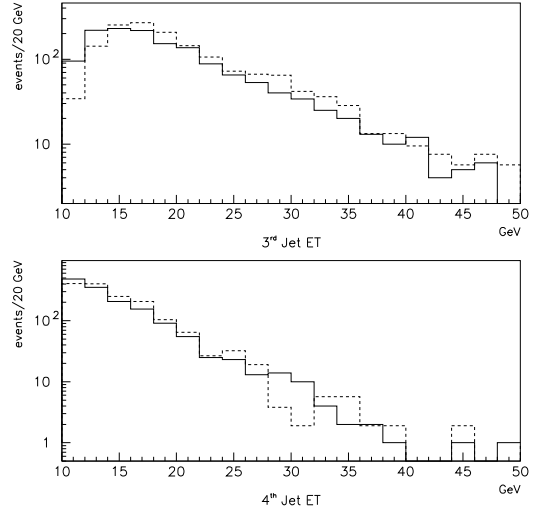,width=3.0in,height=3.0in}}
\caption{ $E_T$ distributions for $3^{rd}$ and $4^{th}$ jet in events with
$\ge 4$ jets starting from VECBOS with 3 jets (dotted lines)
and from VECBOS with 4 jets (solid lines).}
\label{fig-vecbos34}
\end{figure}

\begin{figure}
\vspace*{4pc}
\centerline{\psfig{figure=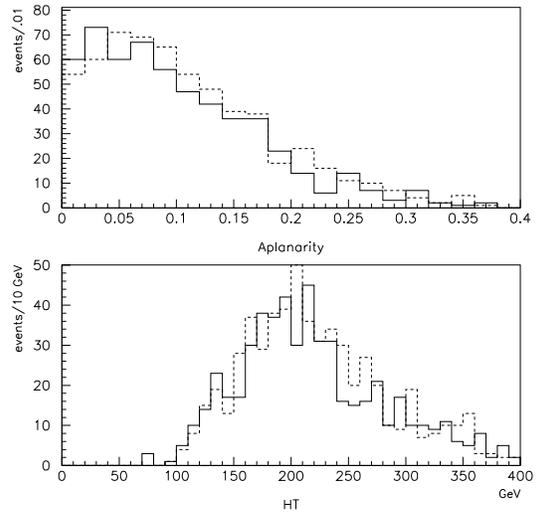,width=3.0in,height=3.0in}}
\caption{${\cal A}$ and $H_T$ distributions for \ttbar\ events with 4
or more jets for top mass 140 GeV. Events generated by ISAJET are in
solid lines, by HERWIG in dotted lines.}
\label{fig-ISA_HWG}
\end{figure}

    Another method to estimate the number of \ttbar\ and \Wjet\ events in the
lepton + jets data is to use event shape information to discriminate
between them. As the ratio of signal/background is small for
jet multiplicities below 4, this analysis concentrates on events with 4
or more jets. Two parameters found to be useful discriminants are
$H_T$, defined as the scalar $E_T$ sum over jets with $|\eta|<2.0$,
and the aplanarity (${\cal A}$), defined as 3/2 times the smallest
eigenvalue of the laboratory 3-momentum tensor of reconstructed objects,
normalized to unit trace. Fig. \ref{fig-aht} shows the distribution
of events in this two variables for the QCD sample, VECBOS Montecarlo,
\ttbar\ events with mass 180 GeV and the data. If we divide the
${\cal A}$, $H_T$ into 4 quadrants using axes ${\cal A}=0.05$ and
$H_T=140$ GeV we see that the QCD sample and the \Wjet\ Montecarlo
sample populate those quadrants more or less equally while the
\ttbar\ events are concentrated in the quadrant with ${\cal A}>.05$
and $H_T>140$ GeV. In table \ref{tab-aht} we give the expected fractions
of \Wjet\ , QCD and \ttbar\ events in each quadrant. To estimate the
uncertainties in the \Wjet\ estimation we generated \Wjet\ events
in two different ways: one was to use VECBOS to generate $W$ + 3 jets events
(at the parton level) and use ISAJET with those events as input to generate
events with 4 or more jets. The other was to use VECBOS to generate
$W$ + 4 jets and then proceed as in the previous case. One can see from
Fig. \ref{fig-vecbos34} that both sets generate quite similar $E_T$
distributions for the 3rd and 4th jet in events with $\ge 4$ jets.
As shown in the
table \ref{tab-aht} the fractions estimated with the two different Montecarlo
samples differ by less than 20\%. The fractions for \ttbar\ were estimated
using ISAJET, comparison of ISAJET with HERWIG showed the differences
to be less than 10\%. Figure \ref{fig-ISA_HWG} illustrates this
for \ttbar\ events generated with ISAJET and HERWIG at a top mass
of 140 GeV.
The number of \ttbar\ events in a given quadrant i ($N_{t \bar t}^i$)
is given by
$$N_{t \bar t}^i=\epsilon_{t \bar t}^i \cdot f_{t \bar t} \cdot N$$
where N is the total number of observed events with
4 or more jets, $\epsilon_{t \bar t}^i$ is the fraction of \ttbar\
expected in quadrant i and $f_{t \bar t}$ is the fraction of N that are
\ttbar\ events.
The number of expected background events is then:
$$N_{bkg}^i=\epsilon_{bkg}^i \cdot (1-f_{t \bar t}) \cdot N$$
where $N_{bkg}^i$ is the number of background events in quadrant i
and $\epsilon_{bkg}^i$ is the expected fraction.
Given N and the $\epsilon $'s one can fit
for $f_{t \bar t}$ using Poisson statistics.
\begin{table}
\Table{|c|c|c|c|c|}{
{}~ & $\epsilon ^1$ & $\epsilon ^2$ & $\epsilon ^3$ & $\epsilon ^4$ \\
 ~&${\cal A}>.05$ & ${\cal A}>.05$ & ${\cal A}<.05$& ${\cal A}<.05$  \\
{}~&$H_T>140$ & $H_T<140$  & $H_T<140$ &$H_T>140$ \\
\hline
VECBOS(4) &
$0.21 \pm .025$ & $ 0.27 \pm .03$ & $0.21 \pm .025$ & $0.31 \pm .03$ \\
VECBOS(3)
& $0.19 \pm .04$ & $ 0.28 \pm .05$ & $0.25 \pm .05$ & $0.28 \pm .05$   \\
QCD
& $0.19 \pm .04$ & $ 0.25 \pm .05$ & $0.28 \pm .05$ & $0.28 \pm .05$   \\
{}~&~&~&~&~ \\
\hline
\ttbar\ 180
& $0.60 \pm .05$ & $ 0.02 \pm .01$ & $0.02 \pm .01$ & $0.36 \pm .04$   \\
\hline
}
\caption{ Fractions in $\geq 4~{\rm jets}$}
\label{tab-aht}
\end{table}
The results of the fit give $f_{t \bar t}=0.32 \pm 0.30$
events in the sample of data with 4 or more jets.
The numbers of observed events (in the 4 quadrants of Fig. 4)
are 4, 1, 3, 4 while the predicted numbers
are 4.1, 2.4, 2.0, 3.5.
There is a large systematic error (35\%) from the choice of partitioning
the ${\cal A}$, $H_T$ plane, this error is estimated by moving the
axes until one event falls into a different quadrant. An additional
20\% systematic error is estimated for the uncertainty in calculating
the fractions for each quadrant. The number of \ttbar\ events estimated
in the sample with 4 or more jets is thus
$N_{t \bar t}=3.8 \pm 3.6 (stat.) \pm 1.5(sys.)$ which leads to a
\ttbar\ production cross section of
$$\sigma _{t \bar t}=8.1 \pm 6.9{\rm (stat.)} \pm 3.8{\rm (sys.)}~pb$$
additional systematic errors in this estimate are the luminosity
(12\%), and the \ttbar\ acceptance (20\%). This value is in good
agreement with that obtained with a totally different method in
section \ref{jetmult}.

The two methods used to estimate $\sigma _{t \bar t}$ can also be used as an
estimate of the number of background events after applying cuts
of ${\cal A}>.05$ and $H_T>140$ GeV which improve the signal/background ratio
by a factor of 3. For method 1 (section \ref{jetmult})
we add the fraction of $W + \ge 4$ jets (predicted
from the fit)
and the fraction non-W events, expected after cuts, to obtain
$N_{bckg}(cut)=1.8 \pm 0.75 \pm 0.40$. From the method used in this
section, the background is given by
$N_{bckg}(cut)=\epsilon_{bckg}^1 \cdot (1-f_{t \bar t}) \cdot N$
which gives $N_{bckg}(cut)=1.7 \pm 0.8 \pm 0.5$. The two ways of
estimating the background share in common only the fraction expected
after ${\cal A}$, $H_T$ cuts. The total amount of background
to the 4 or more jets channel
is estimated in one case by the fit to jet multiplicities, in the other
by a fit to the ${\cal A}$, $H_T$ plane. It is worth emphasizing that
neither method uses the total number of $W$ + $\ge$4 jet events predicted
from Montecarlo.
We can compute
the \ttbar\ cross section after doing a background subtraction (averaging
the background obtained with the 2 methods)
to get $\sigma _{t \bar t}=7.3 \pm 7.2 \pm 3.3~pb$
for $m_{top}=160~-~180$ GeV. The value should be increased by 25\% for
 $m_{top}=140$ GeV.

\section{Conclusion}
    Two different ways of analyzing lepton+jets final states
lead to a similar estimate of the top production cross section
in \ppbar\ interactions at 1.8 TeV center-of-mass energy:
$$\sigma _{t \bar t}=6.4 \pm 9.8 \pm 4.0 ~pb$$
$$\sigma _{t \bar t}=8.1 \pm 6.9 \pm 3.8 ~pb$$
and of the background after ${\cal A}>.05$,
$H_T>140$ GeV cuts:
$$1.8 \pm 0.9,~{\rm and}~1.7 \pm 0.9$$
with 4 events observed.

The data shows a small excess of events above the expected background
but the excess is not statistically significant.
The excess above background corresponds to a
top cross section of $\sigma _{t \bar t}=7.3 \pm 7.2 \pm 3.3~pb$.
These results should be considered preliminary.

We thank the Fermilab Accelerator, Computing and Research Divisions, and
the support staffs at the collaborating insititutions for their contribution
to the success of this experiment. We also acknolwedge the support
provided by the U.S. Department of Energy, the U.S. National Science
foundation, the Commisariat \`{a} L'Energie Atomique in France, the Ministry
for
Atomic Energy in Russia, CNPq in Brazil, the Department of Atomic Energy in
India, Colciencias in Colombia, and CONACyT in Mexico.

\end{document}